# On the interplay between electrical conductivity and Seebeck coefficient in ultra-narrow silicon nanowires

Neophytos Neophytou and Hans Kosina

Institute for Microelectronics, TU Wien, Gußhausstraße 27-29/E360, A-1040 Wien, Austria

e-mail: {neophytou|kosina}@iue.tuwien.ac.at

## Abstract

We analyze the effect of low dimensionality on the electrical conductivity ($\sigma$) and Seebeck coefficient ($S$) in ultra-narrow Si nanowires (NWs) by employing atomistic considerations for the electronic structures and linearized Boltzmann transport theory. We show that changes in the geometrical features of the NWs such as diameter and orientation, mostly affect $\sigma$ and $S$ in two ways: i) the distance of the band edges from the Fermi level ($\eta_F$) changes, and ii) quantum confinement in some cases strongly affect the effective mass of the subbands, which influences the conductivity of the NWs and $\eta_F$. Changes in $\eta_F$ cause exponential changes in $\sigma$, but linear changes in $S$. $S$ seems to be only weakly dependent on the curvature of the bands, the strength of the scattering mechanisms, and the shape of the DOS($E$) function, contrary to current view. Our results indicate that low dimensionality has a stronger influence on $\sigma$ than on $S$ due to the stronger sensitivity of $\sigma$ on $\eta_F$. We identify cases where bandstructure engineering through confinement can improve $\sigma$ without significantly affecting $S$, which can result in power factor improvements.

# I. Introduction

The ability of a material to convert heat into electricity is measured by the dimensionless figure of merit $ZT=\sigma S^2 T/(\kappa_e+\kappa_l)$, where $\sigma$ is the electrical conductivity, $S$ is the Seebeck coefficient, and $\kappa_e$ and $\kappa_l$ are the electronic and lattice parts of the thermal conductivity, respectively. Some of the best thermoelectric materials are based on rare earth or toxic elements and exhibit $ZT \sim 1$. Recent breakthrough experiments on nanostructured thermoelectrics, however, have demonstrated that $\kappa_l$ can be significantly suppressed, offering large improvements in $ZT$ compared to the raw materials' values. Such effects have been observed for 1D nanowires (NWs) [1, 2], 2D thin films, 1D/2D superlattices [3, 4], as well as materials with embedded nanostructuring [5]. More importantly, this has been achieved for common materials such as Si, SiGe and InGaAs. In silicon, although the bulk material has $ZT_{bulk} \sim 0.01$, the $ZT$ of silicon NWs with side lengths scaled down to 10-50nm was experimentally demonstrated to be $ZT_{NW} \sim 0.5$ [1, 2].

On the other hand, it has been suggested by Hicks and Dresselhaus [6] that low dimensionality can be beneficial to the power factor $\sigma S^2$ as well [7]. The sharp features in the low-dimensional density of states, DOS($E$), can improve $S$, as this quantity is proportional to the energy derivative of DOS($E$). This was actually the initial drive towards low dimensional thermoelectrics. Although $S$ and $\sigma$ are inversely proportional, it was suggested that low-dimensional DOS($E$) could potentially improve $S$ without reducing $\sigma$. This effect, however, has not yet been experimentally confirmed because the sharp features in DOS($E$) disappear in the presence of non-idealities.

In this work, we discuss the influence of low-dimensionality on the Seebeck coefficient, and the interplay between the Seebeck coefficient and the electrical conductivity in ultra-narrow NWs of diameters below $D$=12nm. We couple the 20 orbital atomistic sp$^3$d$^5$s*-spin-orbit-coupled (SO) tight-binding (TB) model [8] to linearized Boltzmann transport theory [9, 10, 11]. Our analysis shows that low-dimensionality affects the electronic conductivity stronger than the Seebeck coefficient. The Seebeck coefficient is weakly dependent on the curvature of the bands, the strength of the scattering mechanisms, and even the shape of the DOS($E$) function. It is mostly dependent on the difference of the subband edges from the Fermi level ($\eta_F$). This dependence of $S$ on $\eta_F$ is close to linear, but the dependence of $\sigma$ on $\eta_F$ is exponential. We then indicate bandstructure engineering cases for which $\sigma$ is improved with quantum confinement without significant reductions in $S$, situations which can result in power factor improvements.

## II. Approach

The sp$^3$d$^5$s*-SO tight-binding model [8] accurately captures the electronic structures and inherently includes the effects of quantum confinement. It represents a compromise between computationally expensive fully ab-initio methods, and numerically inexpensive but less accurate effective mass models. Our calculations can include up to 5500 atoms, a challenging, but achievable computational task within this model. We consider infinitely long, cylindrical NWs with hydrogen passivated surfaces [12]. No lattice relaxation is assumed for the NW surfaces.

The electrical conductivity and Seebeck coefficient follow from linearized Boltzmann theory as:

$$\sigma = q_0^2 \int_{E_0}^{\infty} dE \left( -\frac{\partial f_0}{\partial E} \right) \Xi(E), \qquad (1a)$$

$$S = \frac{q_0 k_B}{\sigma} \int_{E_0}^{\infty} dE \left( -\frac{\partial f_0}{\partial E} \right) \Xi(E) \left( \frac{E - E_F}{k_B T} \right), \qquad (1b)$$

where the transport distribution function $\Xi(E)$ is defined as [13, 14]:

$$\begin{aligned} \Xi(E) &= \sum_{k_x, n} v_n^2(k_x) \tau_n(k_x) \delta(E - E_n(k_x)) \\ &= \sum_n v_n^2(E) \tau_n(E) g_{1D}^n(E). \end{aligned} \qquad (2)$$

$v_n(E) = \partial E_n / \hbar \partial k_x$ is the bandstructure velocity, $g_{1D}^n(E) = 1/(2\pi\hbar v_n(E))$ is the density of states for the 1D subbands (per spin), and $\tau_n(k_x)$ is the momentum relaxation time for a state with $k_x$ in subband $n$. For this calculation we use the velocity $v_i(k_x)$ instead of the momentum $k_x$ in the definition of the relaxation time as:

$$\frac{1}{\tau_n(k_x)} = \sum_{m,k_x'} S_{n,m}(k_x, k_x')\left(1 - \frac{v_m(k_x')}{v_n(k_x)}\right) \tag{3}$$

The two are equivalent for parabolic bands, and both are valid approximations resulting from more complicated integral equations as described in [15, 16]. The transition rate $S_{n,m}(k_x, k_x')$ for a carrier in an initial state $k_x$ in subband $n$ to a final state $k_x'$ in subband $m$ is extracted from the electronic dispersions and the atomistically extracted wave form overlaps using Fermi's Golden Rule. The entire procedure is described in detail in Ref. [14].

We include phonon and surface roughness scattering (SRS). For phonon scattering we include all relevant mechanisms of bulk silicon [17]. Although we still employ bulk phonons, this should not affect our conclusions significantly as discussed in Refs [9, 11, 18]. All deformation potential values and phonon energies used are from Ref. [17] with the exceptions of $D_{ODP}^{hole} = 13.24 \times 10^{10}$ eV/m, $D_{ADP}^{hole} = 5.34$ eV, and $D_{ADP}^{electron} = 9.5$ eV from Ref. [9, 11, 18], which are more relevant for NWs.

For SRS we assume a 1D exponential autocorrelation function [19] for the roughness with $\Delta_{rms}$ = 0.48nm and $L_C$ = 1.3nm [20]. The momentum relaxation rate is derived from the shift in the band edges with quantization, as described by Uchida and Tagaki [21] and Fang *et* al. [22]:

$$\frac{1}{\tau_{SRS}^n} = \frac{2\pi}{\hbar}\left(\frac{2\Delta_{rms}^2 L_C}{1 + q^2 L_C^2}\right)\left(\frac{q_0 \Delta E_0}{\Delta D}\right)^2 \sum_m g_{1D}^m(E), \tag{4}$$

where $q = k_x - k_x'$. As discussed in Ref. [21], this is a valid approach for describing SRS in ultra-thin channels of a few nanometers in thickness.

# III. Results and Discussion

To illustrate the diameter dependence of electrical conductivity and Seebeck coefficient, Fig. 1 shows these quantities for the n-type [100] oriented NW vs. diameter for carrier concentration n=$10^{19}$/cm$^3$ (close to where the peak of the power factor appears [23]). The dashed lines indicate the phonon-limited results, whereas the solid lines include phonons and SRS. Figure 1a shows the electrical conductivity. Comparing at the same carrier concentration, the conductivity is degraded by ~4X as the diameter is reduced. The effect of SRS causes an additional ~2X degradation (for the lower diameters). This degradation does not appear in the case of ballistic transport where the ballistic conductance $G$ (normalized by the NW's area in nm$^2$) is almost unchanged as the diameter is reduced (Inset of Fig. 1a) [24].

The Seebeck coefficient in Fig. 1b increases as the diameter is reduced, especially for diameters below $D$=7nm, following the inverse trend compared to conductivity, since these quantities are inversely proportional. The increase is ~70% and can be observed under scattering as well as ballistic conditions (smaller for ballistic). Including SRS causes only a slight additional increase from the phonon-limited result and ballistic results, indicating that $S$ is to first order independent of scattering.

The explanation for these trends originates from the placement of the subband edges in energy with respect to the Fermi level for each NW. The carrier concentration is given by:

$$n_{3D} = \frac{M}{A} \int_{E_0} g_{1D}(E) f(E - E_F) dE, \qquad (5)$$

where $M$ is the number of subbands, $A$ is the normalization cross section area and is $E_F$ the Fermi level. As the area is reduced, the number of subbands $M$ decreases, usually linearly for the thicker NWs such that the ratio $M/A$ remains constant. At some point, only a few or even only one subband participates in transport. Usually for Si at room temperature this happens at $D$<10nm. Further reduction of the NW area will not be linearly compensated by reduction in $M$, and the ratio $M/A$ will increase following $\approx 1/A$ as $M$ approaches closer to 1. To keep the carrier concentration $n_{3D}$ constant, the energy integral has to be reduced, which is achieved when the distance of the subband edges $E_0$ from the Fermi level $\eta_F = E_0 - E_F$ is increased. This is demonstrated in Fig. 2, which compares the position of the Fermi level for n-type [100] NWs with $D$=12nm and $D$=3nm at the same carrier concentration. Figures 2a and 2b show the dispersion relations for the two NWs respectively, for n = $10^{19}$/cm$^3$. The dispersions are shifted to $E_0$=0eV, and the position of the Fermi level is indicated. $\eta_F$ is larger for the $D$=3nm NW. Figure 2c shows the $\eta_F$ for the two NWs vs. carrier concentration. At carrier concentrations from $10^{18}$/cm$^3$ to $10^{20}$/cm$^3$, where the power factor in Si is the highest, the difference in $\eta_F$ between the two NWs is ~40meV. It is also important to note the dependence of the shift of $\eta_F$ on changes of the ratio $M/A$. Using $g_{1D}^n(E) \propto \sqrt{m_{eff}/\tilde{E}}$ (valid for 1D and parabolic bands), where $\tilde{E} = E - E_0$, and $m_{eff}$ being the effective mass of the subband, then:

$$n_{3D} = m_{eff}\left(\frac{M}{A}\right)\int_{E_0}^{\infty} \tilde{E}^{-1/2} f(E - E_F) dE, \qquad (6)$$

The energy integral is an exponential function of $\eta_F$ through the Fermi distribution (under non-degenerate conditions, for $\eta_F > 0$). In order to keep the same carrier concentration under changes in $M/A$ (assuming constant $m_{eff}$) $\eta_F$ changes, but it needs to do so only *logarithmically*. From Eq. 6, therefore, one can observe that to first order, $\Delta\eta_F \propto \ln(M/A)$. The inset of Fig. 2c shows $\eta_F = E_C - E_F$ versus $\ln(A/M)$. We have flipped the numerator and denominator to have the narrowest NWs to the left. Here, for $M$ we use the density of states below a cut-off energy of 0.2eV above the conduction band edge. For larger diameters, $A/M$ and $\ln(A/M)$ are almost constant (right side of Inset), and $\eta_F$ shifts little. Most of the change in $\ln(A/M)$ and $\eta_F$ comes at lower diameters, as expected, where $\eta_F$ follows a more or less linear trend.

When $\eta_F$ increases the conductivity decreases. For a rough qualitative understanding of how the conductivity is affected, we substitute Eq. 2 into Eq. 1a. We assume that $\tau_n(E)$ follows a simple relation (at least for elastic isotropic processes such as acoustic phonon scattering) as:

$$\tau_n(E) \propto \frac{A}{Mg_{1D}^n(E)} \quad (7)$$

which just means that the scattering rates are proportional to the density of states that a carrier can scatter into. Now we substitute these relations in Eqn. 1a (and after performing the summation over the subbands in Eqn. 2):

$$\sigma \propto \int_{E_0} v_n^2(E) \frac{A}{Mg_{1D}^n(E)} \frac{Mg_{1D}^n(E)}{A} \left(-\frac{\partial f(E-E_F)}{\partial E}\right) dE$$
$$= \int_{E_0} v_n^2(E) \left(-\frac{\partial f(E-E_F)}{\partial E}\right) dE \quad (8)$$

Using $g_{1D}^n(E) \propto \sqrt{m_{eff}/\tilde{E}}$ and $v_n(E) \propto \sqrt{\tilde{E}/m_{eff}}$, then:

$$\sigma \propto \int_{E_0}^{\infty} \tilde{E}/m_{eff} \left(-\frac{\partial f(E-E_F)}{\partial E}\right) dE = \frac{1}{m_{eff}} \tilde{F}(\eta_F) \quad (9)$$

where $\tilde{F}(\eta_F)$ is a function of $\eta_F$, independent of bandstructure at first order, and exponentially decreasing with increasing $\eta_F$ (again under non-degenerate conditions, for $\eta_F > 0$). As described earlier, when the NW area is reduced, the ratio $M/A$ increases, forcing $\eta_F$ to also increase logarithmically (as $M$ approaches to 1) in order to keep a constant carrier concentration. This results in a $\approx 1/A$ decrease in the function $\tilde{F}(\eta_F)$ ($\Delta \tilde{F}(\eta_F)$) with cross section area scaling. Following the $\tilde{F}(\eta_F)$ trend, the conductivity decrease ($\Delta\sigma$) also follows $\approx 1/A$ to first order (or $1/D^2$ as indicated in the phonon-limited results in Fig. 1a).

Similarly, the Seebeck coefficient can be shown to follow:

$$S \propto \frac{\int_{E_0}^{\infty} F(\eta_F) \left(\frac{E-E_F}{k_B T}\right) dE}{\int_{E_0}^{\infty} F(\eta_F) dE} \quad (10)$$

where $F(\eta_F) = \tilde{E}\left(-\frac{\partial f(E-E_F)}{\partial E}\right)$. The Seebeck coefficient is therefore at first order independent of bandstructure, and reduces linearly as the energy deviates from the Fermi level as expected ($F(\eta_F)$ is found in both numerator and denominator). As shown in Fig. 1b, at larger NW diameters where $\eta_F$ is small and does not vary significantly, $S$ is constant. As the NW diameter is reduced and $\eta_F$ increases logarithmically, $S$ also increases logarithmically. The magnitude of this logarithmic increase in $S$ with $A$ (or $D$) scaling, however, is smaller compared to the magnitude of the decrease in $\sigma$ which follows $1/A$ (or $1/D^2$).

It is important however, to stress out that the trend presented in Fig. 1 is at constant carrier concentration. Alternatively, the diameter behavior can be presented in terms of constant $\eta_F$. Figure 3 shows the phonon-limited electrical conductivity and Seebeck coefficient vs. the NW diameter for cases: i) constant carrier concentration (same as in Fig. 1), and ii) constant $\eta_F = k_B T$. The behavior under constant $\eta_F$ is different. The electrical conductivity (Fig. 3a) is almost constant (also indicated from Eq. 9), with a slight increase as the diameter is reduced because of the reduction in the available subbands and states the carriers can scatter into. The Seebeck coefficient (Fig. 3b) is also almost constant with diameter, as can be understood from Eq. 10. This is an interesting observation, which shows that it is the distance of the band edges from the Fermi level that controls $S$, whereas the shape of the DOS($E$) does not affect $S$ significantly. Indeed, the DOS($E$) for the $D$=3nm NW shows 1D like behavior, whereas that of the $D$=12nm NW is different, with many more closely packed subband peaks. This is in agreement with the results by Kim *et* al. [25] for 1D, 2D and 3D channels. However, no matter how one presents the diameter dependence, either at constant concentration of constant $\eta_F$, the effect of reducing the diameter from $D$=12nm (almost bulk-like) to $D$=3nm (1D), does not improve the power factor significantly as shown in Fig. 3c. There is only a moderate improvement for smaller diameters if $\eta_F$ is held constant as the diameter is reduced.

There are situations, however, where the effective mass of the subbands reduces as the NW diameter is reduced. As we have shown in our previous works, this is the case for p-type [110] and [111] NWs [11, 26], and at a smaller degree for n-type [110] NWs [27]. This is shown in Fig. 4a and 4b for the p-type [110] $D$=12nm and $D$=3nm NWs, respectively. The curvature of the subbands of the $D$=3nm NW is larger than that of the $D$=12nm, indicating a smaller $m_{eff}$ and DOS compared to the $D$=12nm NW. The reason behind the effective mass reduction with diameter scaling is related to the strongly anisotropic warped shape of the heavy-hole valence band as shown in the inset of Fig. 4a. For bulk materials, the effective mass is determined by the curvature of the $E(k)$ relation along the dashed line passing though the center of the Brillouin zone. Upon confinement, the low-dimensional subbands are formed from energy subbands away from the center (direction of the arrow) as indicated by the solid line in the inset of Fig. 4a, similar to the "particle-in-a-box" quantization picture. Since the heavy-hole is highly anisotropic, the curvature of the bands increases significantly, and the effective mass is reduced. We note that the lines shown represent confinement along the (110) surface and transport along the [110] orientation, which are the relevant orientations for the [110] NW. In the case of n-type NWs, because the conduction band is mostly isotropic along a specific direction, much smaller mass variations are observed [27].

For the p-type [110] case $\eta_F$ will not increase with diameter scaling, in contrast to the n-type [100] NWs in Fig. 2a and 2b. The Fermi level calculated for carrier concentrations p = $10^{19}$/cm$^3$ is almost at the same position for both NWs. Figure 4c shows that $\eta_F$ is almost the same for both NWs in a large range of carrier concentrations, except at very high ones. This is a result of two counter-acting mechanisms: i) as the diameter is reduced, $\eta_F$ tends to increase, but ii) as the effective mass reduces, in order to maintain the same carrier concentration, $\eta_F$ is reduced back. These two counter acting effects finally leave $\eta_F$ almost unchanged. In other words, the increase in $M/A$ in Eq. 6 is compensated by the reduction in $m_{eff}$, and $\eta_F$ remains unchanged. The trend of the conductivity and the Seebeck coefficient as a function of diameter will therefore be the same, at either constant concentration, or constant $\eta_F$.

Figure 5 shows $\sigma$ and $S$ for p-type [110] NWs as a function of the NW diameter at a constant carrier concentration of p = $10^{19}$/cm$^3$. In all cases, $\eta_F$ is very similar, $\eta_F = 0.018 \pm 0.001\ eV$. Phonon-limited (dashed-circled) and phonon- plus SRS-limited (solid-triangle) results are shown. The electrical conductivity in Fig. 5a increases as the diameter is reduced, because the effective mass of the subbands decreases [11]. As indicated in Eq. 9, at constant $\eta_F$ the only contribution to $\sigma$ is from $m_{eff}$. SRS degrades the conductivity for the small diameter NWs, but in this case the improvement due to the reduction in $m_{eff}$ is large enough to compensate for this detrimental effect. The Seebeck coefficient on the other hand in Fig. 5b, does not change significantly with diameter since $\eta_F$ is constant, as can also be deduced from Eq. 10. It follows a slightly reducing trend for most of the diameter range, a reverse trend compared to $\sigma$. At diameters below $D$=5nm, some increase is observed, which is again a result of the trend of $\eta_F$ as shown in the inset of Fig. 5b. Another important observation is that the introduction of SRS in the calculation, only slightly affects $S$. Unlike $\sigma$, $S$ at first order only depends on $\eta_F$ and not on the bandstructure, therefore, the introduction of more scattering mechanisms does not have a significant influence on it. The large increase in $\sigma$ and the almost invariant $S$ diameter trend in this p-type [110] NW case, allows for improvements in the power factor of the channel.

## IV. Conclusions

The interplay between the electrical conductivity ($\sigma$) and the Seebeck coefficient ($S$) in narrow Si NWs of diameters below 12nm is investigated. The sp$^3$d$^5$s$^*$ atomistic tight-binding model and linearized Boltzmann theory are employed. We show that for a specific carrier concentration, as the diameter of the NWs is reduced, the band edges shift further with respect to the Fermi level ($\eta_F$ increases) to first order logarithmically as a function of the NW's cross section area. An increase in $\eta_F$ reduces $\sigma$ exponentially and increases $S$ linearly. Due to the exponential dependence, $\sigma$ is the quantity with the largest influence on the power factor of NWs. As a function of diameter, the logarithmic increase of $\eta_F$, to first order decreases $\sigma$ as $D^2$ and increases $S$ logarithmically. The curvature of the bands, the strength of the scattering mechanisms, and the shape of the DOS($E$) function does not seem to affect $S$ significantly as the diameter changes from $D$=3nm to $D$=12nm in agreement with other reports [25]. We show that in cases where the effective mass of the dispersion becomes lighter with confinement (i.e p-type [110] NWs), $\eta_F$ is less susceptible to NW diameter changes. In such case $\sigma$ increases because of the $m_{eff}$ reduction, whereas $S$ changes only slightly because $\eta_F$ changes only slightly. Improvements in the power factor can in this way be achieved.

## Acknowledgements

This work was supported by the Austrian Climate and Energy Fund, contract No. 825467.

Figure 1:

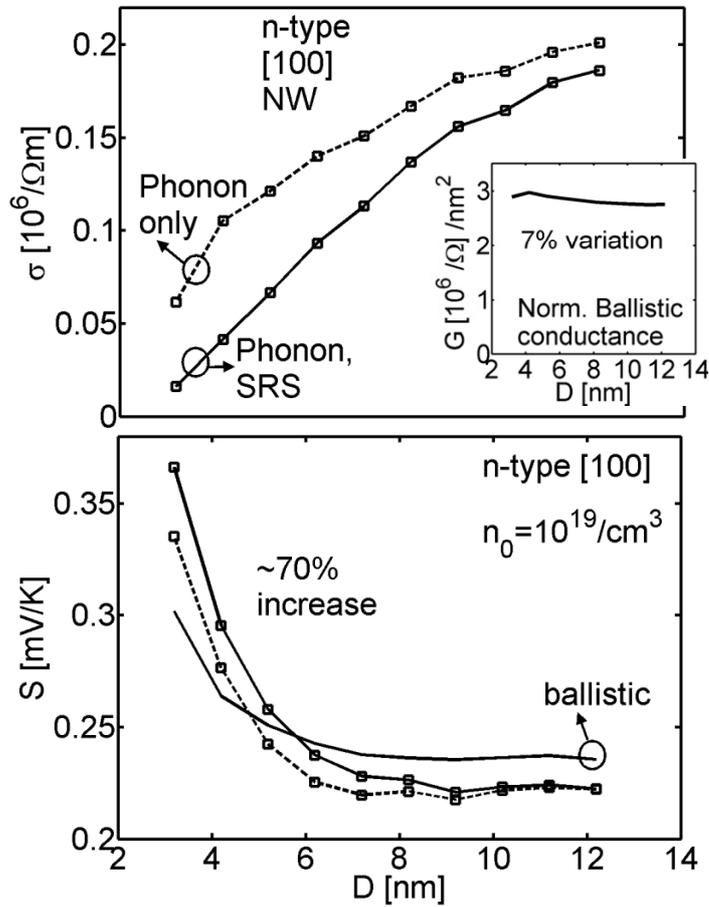

Figure 1 caption:

Electrical conductivity (a), and Seebeck coefficient (b), for n-type NWs in the [100] transport orientation at n = $10^{19}$/cm$^3$ versus diameter. Dashed-squared lines: Only phonon scattering is considered. Solid-squared lines: Phonons and SRS are considered. Solid lines: Ballistic conditions. Inset of (a): The ballistic conductance per unit area (normalized by the nanowire area in nm$^2$) versus diameter.

Figure 2:

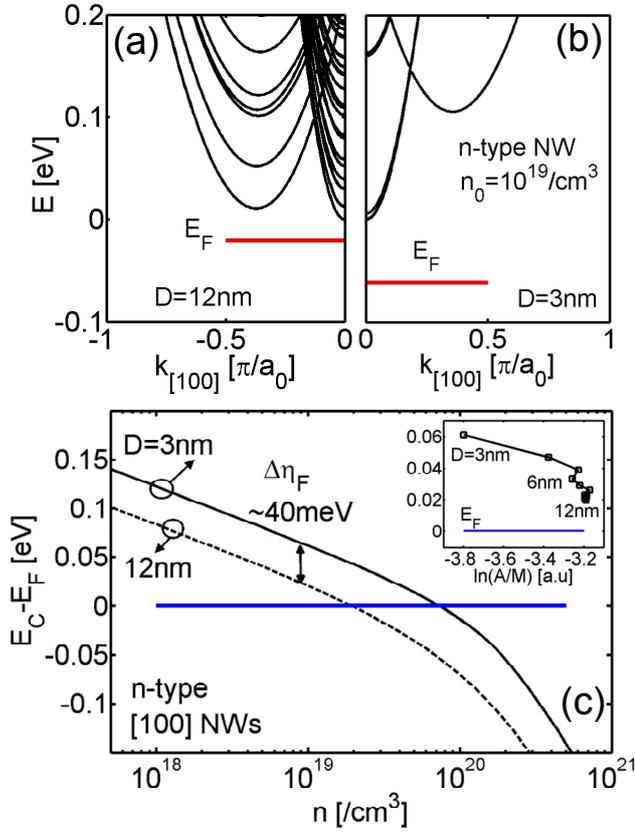

Figure 2 caption:

Electronic structures for n-type [100] NWs: (a) $D$=12nm, (b) $D$=3nm. The position of the Fermi level for carrier concentrations n = $10^{19}$/cm$^3$ is shown for each case. (c) The difference of the dispersion band edges from the Fermi level ($\eta_F$) for the NWs with $D$=12nm (dashed) and $D$=3nm (solid) vs. the carrier concentration. Inset of (c): $\eta_F$ versus ln($A/M$), where $A$ is the NW area. For $M$ we have used the density of states up to 0.2eV from the edge of the conduction band edge.

Figure 3:

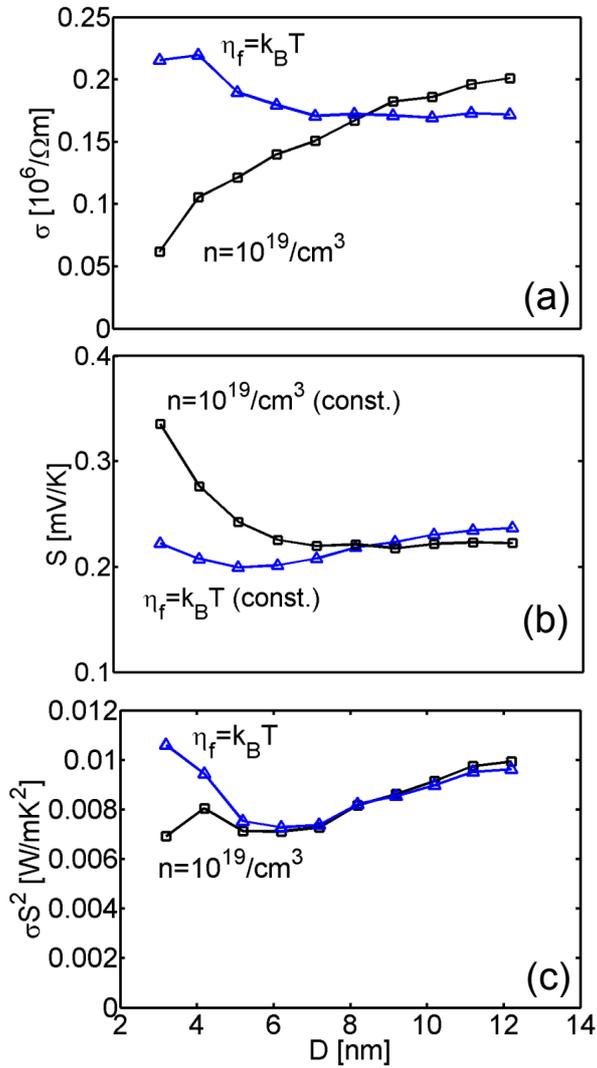

Figure 3 caption:

Thermoelectric coefficients for n-type NWs in [100] transport orientation vs. diameter. Two conditions are shown: i) Constant carrier concentrations n = $10^{19}$/cm$^3$ (squared-black lines). ii) Constant $\eta_F = k_B T$ (triangle-blue lines). (a) The electrical conductivity. (b) The Seebeck coefficient. (c) The power factor.

Figure 4:

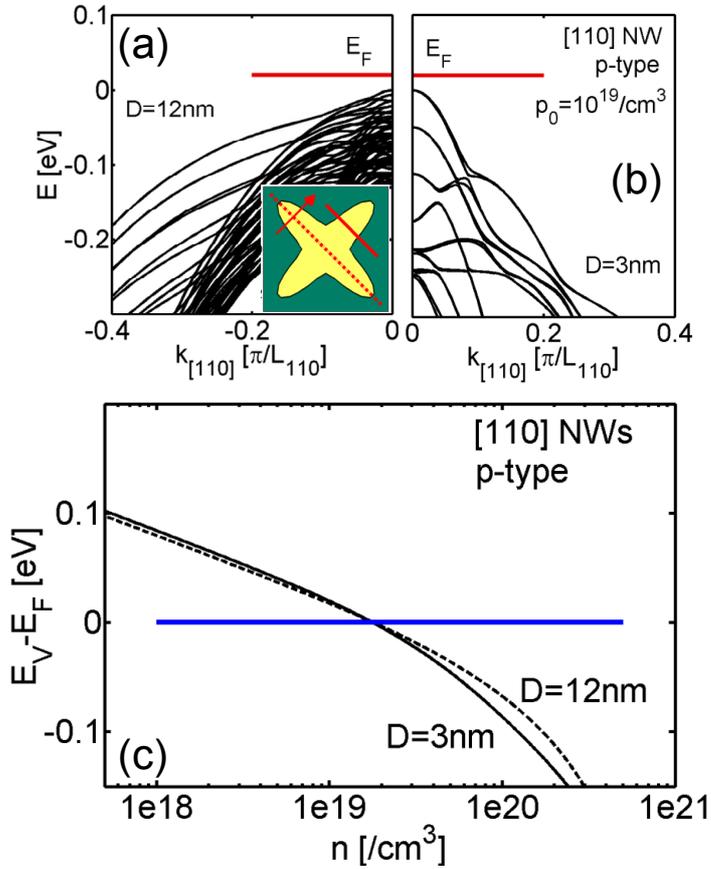

Figure 4 caption:

Electronic structures for p-type [110] NWs: (a) $D$=12nm, (b) $D$=3nm. The position of the Fermi level for carrier concentrations p = $10^{19}$/cm$^3$ is shown for each case. (c) The difference of the dispersion band edges from the Fermi level ($\eta_F$) for the NWs with $D$=12nm (dashed) and $D$=3nm (solid) vs. the carrier concentration. Inset of (a): Schematic of the heavy-hole band of bulk Si. The dotted line indicates the relevant bulk energy bands. Confinement shifts the relevant bands to the direction of the arrow towards the solid line.

Figure 5:

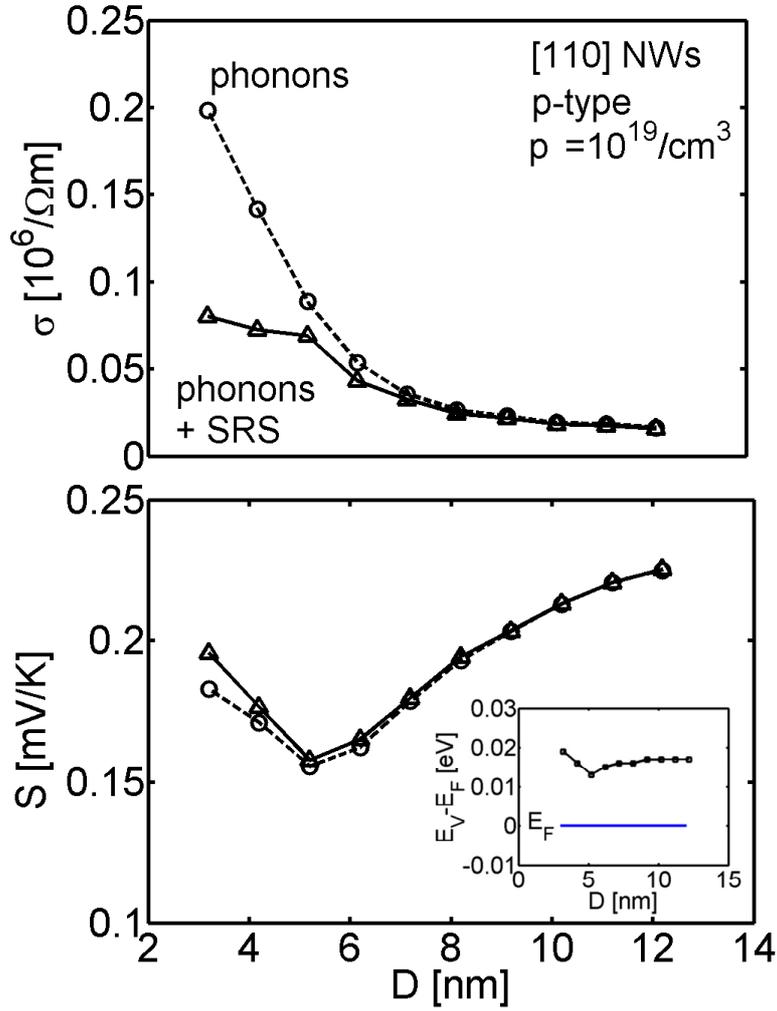

Figure 5 caption:

Thermoelectric coefficients for p-type NWs in the [110] transport orientation vs. diameter at carrier concentrations p = $10^{19}$/cm$^3$. Two conditions are shown: i) Phonon-limited results (dashed-circled lines). ii) Phonon- plus SRS-limited results (solid-squared lines). (a) The electrical conductivity. (b) The Seebeck coefficient. Inset of (b): $\eta_F$ vs. diameter.